\documentstyle[aps,prl]{revtex}
\input epsf
\input{epsf.sty}

\begin{document}
\twocolumn[

\hsize \textwidth\columnwidth\hsize
\csname @twocolumnfalse\endcsname
\setcounter{equation}{0}

\title{Defect-unbinding and the Bose-glass transition in layered superconductors}
\author{C.J. van der Beek$^{1}$, M. Konczykowski$^{1}$, A.V. 
Samoilov$^{2}$, N. Chikumoto$^{3}$, S. Bouffard$^{4}$,
and M.V. Feigel'man$^{5}$}
\address{
$^{{\rm 1}}$Laboratoire des Solides Irradi\'{e}s, Ecole Polytechnique, 91128
Palaiseau, France \\
$^{{\rm 2}}$Condensed Matter Physics 114-36, California Institute of
Technology, Pasadena CA 91125 , U.S.A. \\
$^{{\rm 3}}$Superconductivity Research Laboratory, ISTEC, Minato-ku, 
Tokyo 105, Japan \\
$^{{\rm 4}}$ Centre Interdisciplinaire de Recherche Ions Lasers 
(C.I.R.I.L.), B.P. 5133, 14040 Caen Cedex, France \\
$^{{\rm 5}}$ Landau Institute of Theoretical Physics, Moscow, Russia}

\date{\today}
\maketitle

\begin{abstract}
The low-field Bose--glass transition temperature in heavy-ion irradiated
Bi$_{2}$Sr$_{2}$CaCu$_{2}$O$_{8+\delta}$ increases progressively with 
increasing density of irradiation--induced  
columnar defects, but saturates for densities in excess of $1.5 \times 
10^{9}$ cm$^{-2}$. The maximum Bose-glass temperature
corresponds to that above which diffusion of 
two--dimensional pancake vortices between different vortex lines 
becomes possible, and above which the ``line--like'' 
character of vortices is lost. We develop a description of the 
Bose--glass line that is in excellent quantitative agreement with 
the experimental line obtained for widely different values of track density 
and material parameters.
		
\end{abstract}
] 
\narrowtext

Heavy-ion irradiated (HII) layered  superconductors\cite{Hardy92,Konczykowski93}    
have recently been at the focus of attention, because   
the irradiation--induced amorphous columnar tracks help
overcome the detrimental effects of the high material 
anisotropy \cite{Klein94I,vdBeek95I,Seow96,Doyle96,Zech95}, and partially 
re-establish long-range superconducting 
phase order \cite{Sato97}. In the layered superconductor 
Bi$_{2}$Sr$_{2}$CaCu$_{2}$O$_{8+\delta}$, heavy--ion irradiation increases 
the irreversibility field $B_{irr}(T)$ below which the 
$I(V)$-curve is nonlinear due to vortex pinning on the tracks\cite{Konczykowski95}
to values well above the field $B_{FOT}(T)$ at which the first order 
vortex lattice--to liquid transition field takes place 
in the pristine material \cite{Zeldov95II}.
At inductions $B_{FOT}(T) \ll B \ll B_{irr}(T)$, the irradiated  
superconductor displays the phenomenology of the Bose--glass phase of 
localized vortices\cite{vdBeek95I,Konczykowski95,Nelson92}; 
moreover, the transport properties show
a distinct anisotropy related to the presence of the tracks 
\cite{Klein94I,vdBeek95I,Seow96,Doyle96} that is absent in the pristine 
material \cite{Zech95}, and which suggests that the 
vortices behave as well-defined separate {\em lines},  \em i.e. \rm vortex lines in the 
Bose-glass phase are \em disentangled \rm .

The position of $B_{irr}(T)$ 
and the occurence of flux--line entanglement were shown to be intimately related in 
moderately anisotropic HII superconductors such as 
YBa$_{2}$Cu$_{3}$O$_{7-\delta}$ \cite{Samoilov96}. There, 
$B_{irr}(T) \sim B_{BG}(T)$ corresponds to the  
second order phase transition line between the Bose-glass
and the vortex liquid \cite{Nelson92,Samoilov96,Jiang94}; $B_{BG}(T)$ progressively 
increases with increasing columnar defect density $n_{d}$, to  
an upper limit attained when $n_{d} \approx 1 \times 10^{11} {\rm cm^{-2}}$ 
(corresponding to an ion dose-equivalent ``matching'' field $B_{\phi} \equiv 
\Phi_{0}n_{d} = 2 $ T). Departing from the correspondence of vortex lines 
with the world lines of interacting bosons in two dimensions (2D) 
\cite{Nelson92,Blatter94}, it was argued \cite{Samoilov96} that 
the upper limit of $B_{BG}(T,n_{d})$ corresponds to the field  
beyond which the (entangled) vortex liquid becomes stable 
with respect to the introduction of linear defects. In this 
temperature and field regime, these produce a rather \em weak \rm 
pinning due to intervortex repulsion\cite{Samoilov95,Wengel97} and the averaging effect 
of vortex thermal excursions \cite{Blatter94,Feigelman90}.
The situation in layered superconductors such as
Bi$_{2}$Sr$_{2}$CaCu$_{2}$O$_{8+\delta}$ is apriori 
different. First, the weak coupling between adjacent CuO$_{2}$ bilayers 
(with separation $s \approx $ 1.5 nm)
implies that vortex lines are extremely soft. Over the 
larger part of the first Brillouin zone of the vortex lattice, the contribution of 
the dipole interaction between ``pancake'' vortices in adjacent 
bilayers to the tilt modulus $c_{44}$ is expected to 
exceed that of the line tension $\varepsilon_{1}$, which is determined by the 
interlayer Josephson effect \cite{Glazman91,Blatter96}. 
Flux lines are then better described as stacks of pancakes 
and the analogy with the 2D boson system is no longer valid \cite{Koshelev96}.  
Another consequence is that pinning near $B_{irr}(T)$ in 
HII Bi$_{2}$Sr$_{2}$CaCu$_{2}$O$_{8+\delta}$ 
is not weak: at fields $B \lesssim \frac{1}{6}B_{\phi}$ pancake vortices 
gain maximum free energy by remaining localized on the columnar defects,
even in the vortex liquid phase \cite{vdBeek2000}.
Nevertheless, a well--defined transition from very slow nonlinear vortex 
dynamics in the Bose--glass \cite{vdBeek95I,Konczykowski95}, 
to Ohmic response in the vortex liquid \cite{Seow96,Doyle96}
does exist near the irreversibility line in Bi$_{2}$Sr$_{2}$CaCu$_{2}$O$_{8+\delta}$.

In order to cast light on the mechanism of this transition,
we have measured $B_{irr}(T)$ in HII Bi$_{2}$Sr$_{2}$CaCu$_{2}$O$_{8+\delta}$
for widely varying track density and oxygen content.
The latter determines the values of the anisotropy parameter $\gamma$ and 
the penetration depth $\lambda_{ab}( T )$ \cite{Li96}: both increase 
as one decreases the oxygen content towards optimal doping.  
It turns out that the phenomenological behavior of $B_{irr}(T)$ at low 
fields is rather similar to that found in YBa$_{2}$Cu$_{3}$O$_{7-\delta}$.
$B_{irr}(T)$ increases progressively towards higher values with 
increasing defect density, but saturates for $n_{d} \gtrsim 1.5 \times 
10^{9}$ cm$^{-2}$, or $B_{\phi} \gtrsim 30$ mT. 
For higher matching fields, the low--field portion ($B \lesssim \frac{1}{6}B_{\phi}$)
of $B_{irr}(T)$ adopts a strictly exponential temperature dependence;   
although $T_{irr}(B)/T_{c}$ strongly increases when the value of 
$\lambda_{ab}$ decreases, we will see that this increase is not due to the increase 
of the pinning energy.

Bi$_{2}$Sr$_{2}$CaCu$_{2}$O$_{8}$ crystals were  grown at the University 
of Tokyo using the travelling-solvent floating-zone method, and then
postannealed at either 800$^{\circ}$C or 500$^{\circ}$ C. This 
produces  $T_{c}$'s of 89 K (optimal doping) and 82 K (overdoped) 
respectively \cite{Li96}. The crystals were cut to rectangles of size 500 
(l) $\times$ 400 (w) $\times$ 20 (t) $\mu$m$^{3}$, and
irradiated with varying fluences of 5.8 GeV Pb ions at the Grand Acc\'{e}l\'{e}rateur
National d'Ions Lourds (GANIL) at Caen, France. The ion beam was directed parallel 
to the sample $c$-axis. Each ion impact created an amorphous columnar track of radius 
3.5 nm traversing the sample along its entire thickness. Samples were prepared 
with track densities $1\times 10^{9}$  cm$^{-2} < n_{d} < 2\times 
10^{11}$ cm$^{-2}$, corresponding 
to 20 mT $\leq B_{\phi} \leq$  4 T. The irradiation 
caused $T_{c}$ to decrease: $\partial T_{c}/\partial n_{d} =  
-3.6\times 10^{11}$ Kcm$^{2}$ or $\partial T_{c}/\partial B_{\phi} = 
-1.8$ KT$^{-1}$.

Subsequent measurements were performed using 
the Local Hall Probe Magnetometer in AC mode \cite{vdBeek95I,Gilchrist93}.
A small ac field of amplitude $h_{ac} = 1$ G
and frequency $f = 
7.75$ Hz is applied parallel to the sample $c$--axis, colinearly with the DC field  
used to create the vortices. The ac field leads to a periodic electric field gradient 
of magnitude $ \sim 2\pi\mu_{0}h_{ac}f$ 
across the sample. Using a miniature Hall probe one measures the RMS 
induction $B_{ac}(f,T)$ at the center of the sample top surface, 
which is simply related to the sample screening current  
\cite{Gilchrist93,vdBeek95II}.

\begin{figure}
\centerline{\epsfxsize 7cm \epsfbox{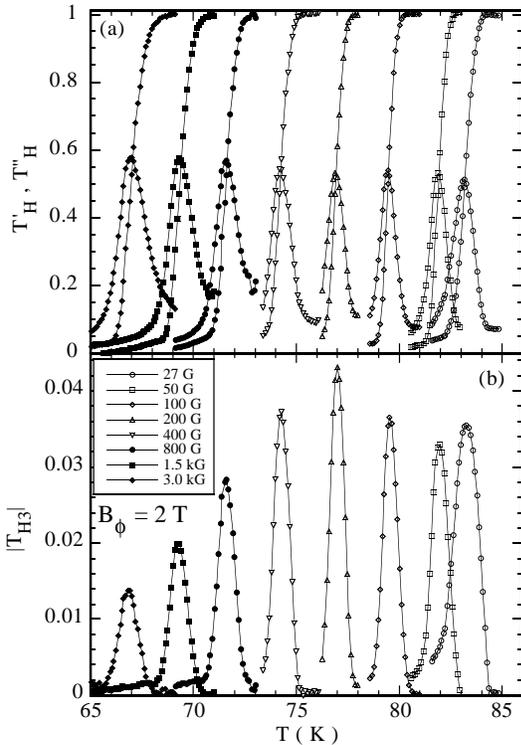}}

\caption{\label{fig:THTH3}
Fundamental (a) and third harmonic (b) transmittivity of the 
Bi$_{2}$Sr$_{2}$CaCu$_{2}$O$_{8}$ single crystal with matching field 
$B_{\phi} = 2$ T, for DC fields (applied parallel to the $c$--axis and to the 
columnar defects) between 27 G and 3.0 kG.}
\end{figure}

Figure~\ref{fig:THTH3} shows the fundamental and 
third harmonic transmittivities, defined as $T_{H}\equiv 
[B_{ac}(f,T) - B_{ac}(f,T\! \ll \!T_{c})]/B$ and $T_{H3}\equiv 
B_{ac}(3f,T)/B$ respectively, with

\begin{figure}
	\centerline{\epsfxsize 7cm \epsfbox{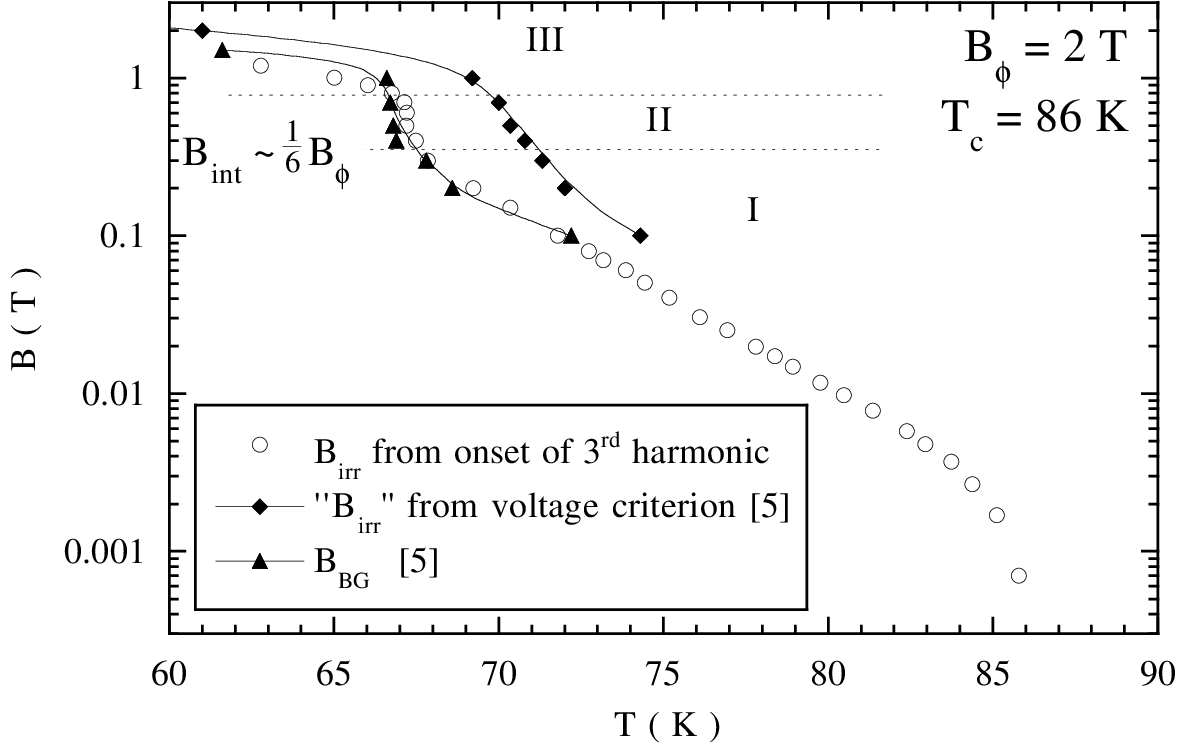}}
	\caption{\label{fig:compare_to_Seow}
	The field $B_{irr}(T)$ below which a third harmonic transmittivity signal can be 
observed in an optimally doped HII Bi$_{2}$Sr$_{2}$CaCu$_{2}$O$_{8}$ single 
crystal ( $B_{\phi} = 2$ T) ($\circ$). Filled triangles denote the Bose-glass transition 
line determined from the power--law scaling of the resistivity, $\rho \propto (T - 
T_{BG})^{s}$ of a crystal with the same $T_{c}$ and $B_{\phi}$ \protect\cite{Seow96};
filled diamonds denote the ``irreversibility line'' determined from the 
onset of a measureable resistivity \protect\onlinecite{Seow96}.}
	
	\end{figure}

\noindent   
$B \equiv B_{ac}(f,T \! \gg \! T_{c})-B_{ac}(f,T  \! \ll \! T_{c})$ \cite{Gilchrist93}, 
measured for the optimally doped crystal with $B_{\phi} = 2 $ T. The presence of a 
third harmonic response $|T_{H3}|$ implies the non--linearity of the sample's
$I(V)$--characteristic. The fields $B_{irr}(T)$, or temperatures $T_{irr}(B)$, 
below which the third harmonic signal can be first observed upon cooling are plotted in 
Fig.~\ref{fig:compare_to_Seow}. At $B_{irr}$, the working point enters the nonlinear part of the 
sample $I(V)$ curve at an electric field of the order of $10^{-7}$ Vm$^{-1}$; this corresponds 
to a voltage drop of $\sim $ 50 pV across the sample. At such low 
voltages, the measurement of the ac screening current may be 
sensitivity--limited \cite{vdBeek95I,vdBeek95II}. In practice, the use 
of $h_{ac} = 1$ Oe means that the minimum measurable current density $j_{min} \approx 5 \times 
10^{2}$ Am$^{-2}$, comparable to a transport current of 0.1 mA. 
The coincidence of our $B_{irr}(T)$ with the $B_{BG}$--data determined by Seow {\em et al.} 
\cite{Seow96} indicates that for all practical purposes $B_{irr}$ is a good 
approximation of the Bose-glass transition field  (Fig.~\ref{fig:compare_to_Seow}).

\begin{figure}
\centerline{\epsfxsize 7cm \epsfbox{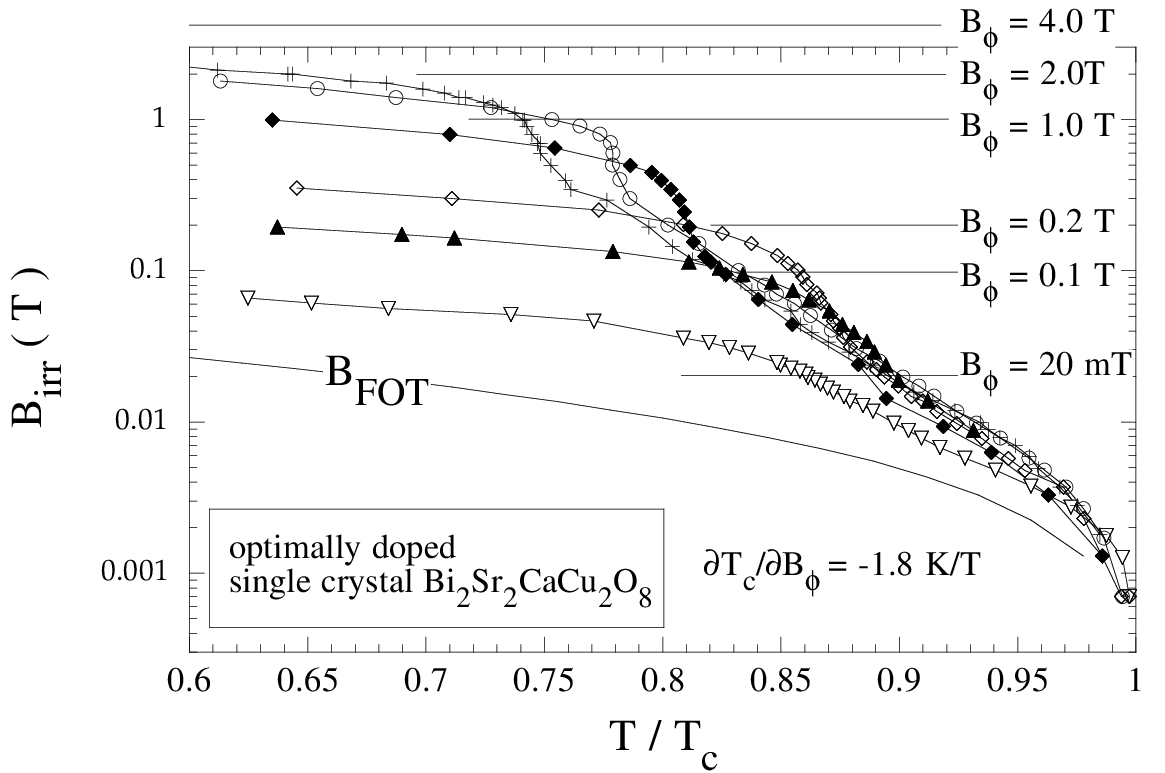}}
\caption{\label{fig:IRLcompl} $B_{irr}(T)$ for optimally doped 
Bi$_{2}$Sr$_{2}$CaCu$_{2}$O$_{8}$  crystals with $B_{\phi}$
between 0 and 4 T. The drawn line indicates the first order phase transition 
$B_{FOT}$ in a pristine crystal.}
\end{figure}

The evolution of $B_{irr}(T)$ with $B_{\phi}$ is plotted in 
Fig.~\ref{fig:IRLcompl}. For $B_{\phi} \lesssim B_{\phi}^{min} 
= 30$ mT, $B_{irr}(T)$  increases monotonically with increasing $B_{\phi}$ 
\cite{Khaykovich98}. As long as $B_{irr}(T) < B_{\phi}$, it depends exponentially 
on temperature, while for $B_{irr} > B_{\phi}$, 
$B_{irr} \propto (1- T/T_{c})$\cite{vdBeek95II}. For large $B_{\phi} 
\gtrsim  B_{\phi}^{min}$, 
one can distinguish \em three \rm distinct sections of 
$B_{irr}(T)$. At all but the very lowest fields, $B_{irr}$ 
again depends exponentially on $T$, but with the specificity that it is
\em independent of \rm $B_{\phi}$. In this regime [(I) in 
Fig.~\ref{fig:compare_to_Seow}], there thus exists a an \em upper limit \rm of $B_{BG}(T)$ in 
Bi$_{2}$Sr$_{2}$CaCu$_{2}$O$_{8}$, as is the case in 
YBa$_{2}$Cu$_{3}$O$_{7-\delta}$ \cite{Samoilov96}. At $B_{irr} \equiv 
B_{int} \sim \frac{1}{6}B_{\phi}$, the exponential increase abruptly changes into a nearly 
vertical rise (regime II);  $B_{int}$ is the field at which intervortex repulsion start to 
determine the pinned vortex configuration\cite{vdBeek2000}. This transition is also
manifest in the vortex liquid phase as that at which a ``recoupling''
transition was measured using Josephson Plasma Resonance (JPR) \cite{Sato97}.
As $B_{irr}$ increases to a sizeable (but not constant) fraction of 
$B_{\phi}$, a weak temperature dependence  is once 

\begin{figure}
\centerline{\epsfxsize 7cm \epsfbox{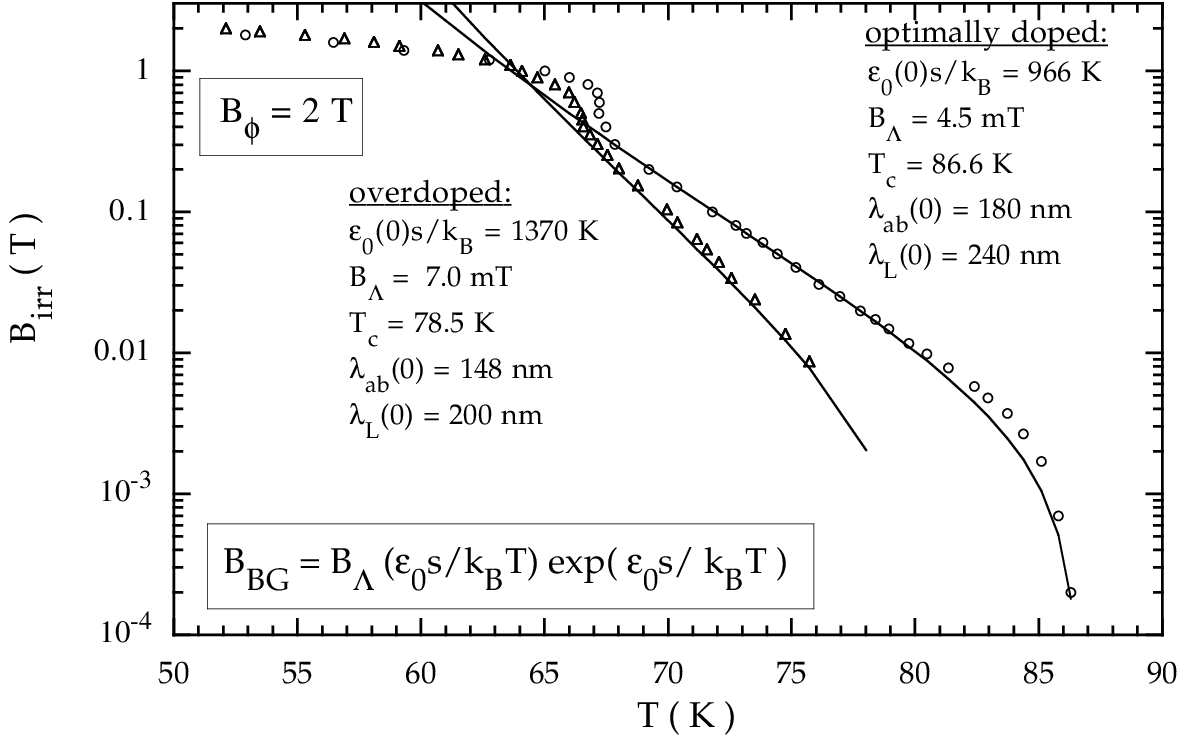}}

\caption{\label{fig:2Tfits} $B_{irr}(T)$ for an optimally doped 
($\circ$ , $T_{c} = 86.3$ K after irradiation) and a lightly 
overdoped ($\triangle$, $T_{c} = 78.5$ K) Bi$_{2}$Sr$_{2}$CaCu$_{2}$O$_{8}$ crystal, both 
with $B_{\phi} = 2$ T. Drawn lines indicate 
fits to Eq.~(\protect\ref{eq:BBG}), with parameter values as 
indicated.}
\end{figure}

\noindent    
again adopted, 
$B_{irr} \propto (1-T/T_{c})^{\alpha}$ with $\alpha \gtrsim 1$ 
(regime III). The behavior of the irreversibility line for different oxygen content 
is shown in Fig.~\ref{fig:2Tfits}. 
The exponential decrease at high $T$ is steeper for the overdoped crystal, 
which has the smaller $\lambda_{ab}$ and $\gamma$, and therefore the 
larger condensation energy $\varepsilon_{0}/4\pi\xi^{2}$, and the stronger intervortex-- and 
vortex-defect interaction (both are proportional to the typical vortex energy scale
$\varepsilon_{0} = \Phi_{0}^{2}/4\pi\mu_{0}\lambda_{ab}^{2}$; $\xi$ is 
the coherence length).

In order to describe the low--field exponential temperature dependence 
of $B_{irr}(T)$, we exploit the fact that in this regime (I) 
vortex interactions are irrelevant to the column occupation  \cite{vdBeek2000}. 
In other words, each vortex line can become localized on an appropriate defect site. 
If a sufficient number of columns is available to every line
(\em i.e. \rm at large $B_{\phi}$), 
extra free energy can be gained by the redistribution of pancakes 
constituting a given line over different columns.  At low 
$B_{\phi} \lesssim 0.5$ T this entropy gain is insufficient to
balance the loss in vortex interaction  energy, and pancakes 
belonging to the same line remain aligned on the same columnar defect. 
In either case, all pancakes are localized on a column.
The superposition of the exponential portions of $B_{irr}(T)$ for \em all \rm matching fields
 30 mT $ < B_{\phi} < 4$ T implies that \em at \rm $B_{irr}$
the  redistribution over different column sites is irrelevant
for the pancake delocalization mechanism ---  \em i.e. \rm at and  
below $B_{irr}(T)$, pancakes belonging to the same vortex line are well aligned 
on the same site, belonging to a set of allowed ``columnar-defect'' 
sites. Then, we no longer need to consider other positions in the ``intercolumn space'' 
in our model description, which becomes that of a ``discrete'' 
superconductor. Since the allowed sites are more or less equivalent, the 
circumstance that they are in fact columnar defect sites becomes 
immaterial: namely, the free energy of all vortices is lowered by 
approximately the same amount. 
The low--field vortex state therefore does not differ 
fundamentally from that in the \em unirradiated \rm crystal. Only,
the vortex confinement in the defect potential, which plays the role of the
 ``substrate potential'' of Ref.~\cite{Dodgson99}, inhibits thermal
 line wandering and the elastic relaxation of the vortex lattice.

\noindent 

The main thermal excitations in this situation are 
expected to be small defects in the vortex lattice. 
In a Josephson--coupled layered superconductor, 
these amount to bound pancake vacancy--interstitial pairs within the 
same layer, {\em i.e.} the ``quartets'' of Ref.~\cite{Feigelman90II}. In the HII layered 
superconductor, such a pair corresponds to the ``exchange'' of one or 
more pancakes between two sites.  The energy of the quartets is \cite{Feigelman90II}

\begin{equation}
\varepsilon_{q} \approx  4  c_{66}a_{0}^{2}s  \left( R / \Lambda \right)^{2} 
\approx \varepsilon_{0} s  \left( R / \Lambda \right)^{2}   ,
\hspace{5mm} (R \ll \Lambda)
\label{eq:epsilon_q}
\end{equation}

\noindent with $c_{66}$ the vortex lattice shear modulus, 
$a_{0} = (\Phi_{0}/B)^{1/2}$ the vortex spacing,
$R$ the distance between a bound vacancy and interstitial, 
$\Lambda = [\lambda_{ab}^{-1} + (\gamma s)^{-1}]^{-1}$ the generalized penetration depth 
taking into account both magnetic and Josephson coupling,  and $\gamma s$ the 
Josephson length \cite{Glazman91,Feigelman90II}. 
It was shown in Ref.~\cite{Feigelman90II} that the glass transition in a layered superconductor 
corresponds to the pair--unbinding transition. 
Correspondingly, the Bose glass transition is the unbinding temperature of the 
dislocation pairs in the ``discrete'' superconductor, \em i.e.\rm   the 
temperature above which pancakes can \em diffuse \rm  from line to line; it 
can be estimated as $ k_{B}T_{BG} = \varepsilon_{q}( R_{l} )$, 
where $R_{l} \approx n_{l}^{-1/2}$ and $n_{l}$ is the equilibrium
density of free dislocation pairs in the  
vortex liquid. Taking into account that only small pairs of 
size $a_{0}$ ( \em i.e. \rm vacancies / interstitials) matter, 
one has $n \approx  a_{0}^{-2} \exp(- \varepsilon_{0}s / k_{B}T)$; 
the activation energy $\varepsilon_{0}s$ is larger than that in the 
unirradiated superconductor because of the lack of lattice relaxation
around the pair. 
Gathering terms, $ k_{B}T_{BG} = \varepsilon_{0} s  (a_{0}/\Lambda )^{2} 
\exp ( \varepsilon_{0}s/k_{B}T_{BG} ) $ or 

\begin{equation}
B_{BG} = B_{\Lambda} \left( \frac{\varepsilon_{0}s}{k_{B}T} \right) 
\exp\left(\frac{ \varepsilon_{0}s}{k_{B}T}\right) ,
\hspace{2mm} ( B_{\Lambda}\ll B \ll  B_{\phi})
\label{eq:BBG}
\end{equation}

\noindent with $B_{\Lambda} = \Phi_{0}/\Lambda^{2}$, $\varepsilon_{0}$ 
and $\Lambda$ to be evaluated at $T_{BG}$. The Bose--glass 
transition line does not depend on the details of the columnar defect 
potential such as pinning energy, column radius, or matching field. 
In Fig.\ref{fig:2Tfits}, we compare Eq.~(\ref{eq:BBG}) to the 
experimental results for Bi$_{2}$Sr$_{2}$CaCu$_{2}$O$_{8}$ 
with different oxygen content. We obtain excellent \em quantitative 
\rm agreement using $\lambda_{ab}$--values from the literature 
\cite{Li96} and the single free parameter $\gamma$ (contained only in 
$B_{\Lambda}$). The values  $\gamma = 360$ and 550 found for the
overdoped and the optimally doped material respectively are well 
within accepted experimental limits. Thus, Eq.~(\ref{eq:BBG}) reproduces 
the correct field, temperature, 
$B_{\phi}$, and doping--dependence of the Bose-glass delocalization line.

The Bose-glass transition separates the low--$T$ phase in 
which pancake vortices can wander between columnar defects, 
but always remain bound to the same site, or vortex line, from the 
high $T$--phase in which diffusion of pancakes between sites (lines) is possible. 
Below $T_{BG}$, individual pancakes (``defect pairs'') cannot provide flux transport 
at low currents; vortex lines can only move as a whole, leading to 
the line--like behavior observed  in 
Ref.~\cite{Klein94I,vdBeek95I,Seow96,Doyle96,Zech95}. 
Oppositely, the pancake diffusion above $T_{BG}$ not only
implies an Ohmic resistivity \cite{Seow96,Doyle96}, but also that the linear 
nature of the vortex lines should no longer be apparent in the angular dependence 
of the transport properties. The intersite diffusion of pancakes (or ``pancake--exchange'') 
means that at $T > T_{BG}$ vortex lines are effectively entangled, 
on a scale ${\mathit l} = s \exp(\varepsilon_{0}s/k_{B}T)$. The upper limit 
of the Bose-glass transition in layered 
superconductors is thus analogous to that in moderately anisotropic 
compounds in that it represents the boundary at which vortices become 
delocalized into an entangled flux liquid and where the angular 
dependence of the columnar defects' contribution to the resistivity 
vanishes.


The meaning of $B_{\Lambda}$ becomes apparent from 
Eq.~(\ref{eq:epsilon_q}), where $\Lambda$ appears as the typical 
interaction distance between small dislocation pairs. For 
separations $\gg \Lambda$, {\em i.e.} for matching fields $B_{\phi} 
\ll B_{\Lambda}$, the formation of quartets demands a (shear) 
energy cost that is much greater than either $\varepsilon_{0}s$ 
or $k_{B}T$. The excitation of a pancake
onto another site is then unlikely even in the vortex liquid.
Delocalization will occur as it does in a 
moderately anisotropic superconductor: thermal wandering of the 
vortices into the intercolumn space lowers their binding energy 
until they can be liberated from the columns by thermal activation \cite{vdBeek95I}. 
Similarly, in regime II ($B > \frac{1}{6}B_{\phi}$), where
the number of available tracks is insufficient not because of the intercolumn 
distance but because of track occupation by other vortices, 
delocalization is initiated by pancake wandering into the intercolumn 
space. Then, $B_{irr}(T)$ \em increases \rm with decreasing $B_{\phi}$ 
because the number of available sites decreases.
The \em experimentally \rm found condition  $B < \frac{1}{6}B_{\phi}$ 
on the validity of Eq.~(\ref{eq:BBG})
\em also \rm describes its validity range at 
\em low \rm $B_{\phi}$: $B_{irr}$ becomes ion--dose dependent when 
$B_{\phi} < B_{\phi}^{min} = 6 B_{\Lambda}$, in excellent agreement with
the experimental values  $B_{\phi}^{min} \sim 30$ mT and 
$B_{\Lambda} = 7.0$ mT (4.5 mT) found for overdoped 
(optimally doped) Bi$_{2}$Sr$_{2}$CaCu$_{2}$O$_{8}$ 
respectively. 

Summarizing, in the regime of individual vortex line 
pinning by columnar defects in a layered superconductor, the
upper limit of the Bose--glass transition corresponds to the onset of 
pancake vortex diffusion between different lines and flux-line 
entanglement. However, in both the glass and the vortex liquid phases,
\em only \rm columnar defect sites are available to the pancakes;  
hence, a modelisation in terms of a  ``discrete superconductor''
is appropriate. The onset of diffusion in this ``discrete superconductor''
presents a realization of the glass transition by unbinding of small 
defect pairs proposed in Ref.~\cite{Feigelman90II}; in the original continuous problem of  
an unirradiated layered superconductor, the defect--unbinding mechanism 
is masked by strong Gaussian thermal fluctuations \cite{Glazman91}.

We thank V.B. Geshkenbein, P.H. Kes, A.E. Koshelev, P. LeDoussal, and V.M. Vinokur
for stimulating discussions. The work of M.V.F. was supported by DGA grant No. 
94-1189.

\end{document}